\documentclass[nouppercase]{ifmbe}

\usepackage[cmex10]{amsmath}
\usepackage{amssymb}
\usepackage{lineno}
\usepackage{epstopdf}

\title{Prior Variances and Depth Un-Biased Estimators in EEG Focal Source Imaging}

\affiliation{Institute for Computational and Applied Mathematics,
University of M\"unster, M\"unster, Germany }{SECONDAFF}
\affiliation{Institute for Biomagnetism and Biosignalanalysis,
University of M\"unster, M\"unster, Germany }{FIRSTAFF}
\affiliation{Department of Electrical and Electronic Engineering,
Imperial College London, London, UK }{THA} \affiliation{Department
of Mathematics, University of Auckland, Auckland, New Zealand }{FOA}

\author{A. Koulouri}{SECONDAFF}
\author{V. Rimpil\"ainen}{FIRSTAFF}
\author{M. Brookes}{THA}
\author{J. P. Kaipio}{FOA}

\begin{document}

\maketitle

\begin{abstract}
In electroencephalography (EEG) source imaging, the inverse source
estimates are depth biased in such a way that their maxima are often
close to the sensors. This depth bias can be quantified by
inspecting the statistics (mean and covariance) of these estimates.
 In this paper, we find weighting factors within a Bayesian framework
for the used $\ell_1/\ell_2$ sparsity prior that the resulting
maximum a posterior (MAP) estimates do not favour any particular
source location. Due to the lack of an analytical expression for the
MAP estimate when this sparsity prior is used, we solve the weights
indirectly. First, we calculate the Gaussian prior variances that
lead to depth un-biased maximum a posterior (MAP) estimates.
Subsequently, we approximate the corresponding weight factors in the
sparsity prior based on the solved Gaussian prior variances.
Finally, we reconstruct focal source configurations using the
sparsity prior with the proposed weights and two other commonly used
choices of weights that can be found in literature.

\end{abstract}



\section{Introduction}
In EEG focal source imaging, the goal is to estimate the focal
neural activity that arises, for example, during
an epileptic seizure using scalp potentials. 
Based on the distributed source modelling \cite{Baillet2001a}, the
mapping that connects the dipole moments of $n$ potential source
locations to $m$ scalp-potential measurements can be
written as
\begin{equation}\label{eq:obm2}
v = Kd + \xi,
\end{equation}
where $v \in \mathbb{R}^m$, $K \in \mathbb{R}^{m \times kn}$ ($m\ll
kn$) is the lead field matrix, $k$ is the dimension of the problem
(2D or 3D), $d\in \mathbb{R}^{kn}$ is the distributed dipole source
configuration and $\xi \sim \mathcal{N}(0,\Gamma_\xi)$ is the
measurement noise.

The ill-posedness of the associated inverse problem requires the use
of prior information to obtain stable estimates. One way to solve
the problem is to find the estimate of the under-determined linear
system that has the minimum norm \cite{Hamalainen1994}. However, the
minimum norm estimate (MNE) has the property that its maxima can lie
only close to the sensors, because
the measured scalp potentials  can be generated from superficial
source configurations with less power than from deep source
configurations \cite{Fuchs1999}. Similar source reconstructions can
also be obtained with $\ell_2$-norm priors. Even if $\ell_1$-norm
priors are employed the solution consists of several scattered
superficial sources \cite{Burger2013a}.

To reduce the depth bias several (often heuristic) approaches have
been suggested
\cite{Kohler1996,Pascual-Marqui1994,Fuchs1999,Wagner2000,Soler2007}.
The most common approaches are to weight all the sources in the
penalty term with the norm of the corresponding column of the lead
field matrix \cite{Kohler1996,Lin2006}
or the diagonal elements of the model
resolution matrix 
\cite{Pascual-Marqui,Haufe2008b}. Another approach is to use the
Bayesian hierarchical modelling \cite{Lucka2012}. 

In this paper, our aim is to find, within a Bayesian framework,
weights for our sparsity prior such that the resulting posterior
estimates do not favor any particular source location or component.
Because there is no analytical expression for the MAP estimate when
sparsity priors are employed, we propose to solve the weights
indirectly. We first quantify the depth bias of the maximum a
posterior (MAP) estimates when an i.i.d. Gaussian prior is employed
by inspecting the statistics of the MAP estimates. Next, we
calculate the Gaussian prior variances that ensure depth un-biased
solutions by equalizing the variances in the covariance matrix of
the MAP estimates. Finally, we approximate the corresponding
weighting factors in the sparsity prior using the solved Gaussian
prior variances. We demonstrate the feasibility of our approach by
simulating focal brain activity with finite element (FE)
simulations. In the reconstructions, we employ the weighted
$\ell_1/\ell_2$ sparsity prior and we compare the results obtained
using our proposed weights with the reconstructions based on two
other commonly used choices of depth weights.

\section{Theory}

\subsection{Bayesian Inversion}

In the Bayesian framework, the inverse solution is the posterior
density of the Bayes formula
\begin{equation}\label{post}
\pi(d|v) \propto \pi(v|d)\pi(d),
\end{equation}
where $\pi(v|d)$ is the likelihood and $\pi(d)$ the prior. From
Equation (\ref{eq:obm2}), the likelihood is
\begin{equation}\label{lik1}
\pi(v|d) \propto
\exp\Big(-\frac{1}{2}(v-Kd)^\mathrm{T}\Gamma_\xi^{-1}(v-Kd)\Big).
\end{equation}
The MAP estimate of the reconstructions is \cite{Kaipio2004},
\begin{equation}\label{eq:MAP}
\hat{d} :=\min_{d\;\in\;\mathbb{R}^{kn}} \|L_\xi(Kd-v)\|_2^2 - 2\ln
\pi(d),
\end{equation}
where $L_{\xi}$ comes from the Cholesky factorization of
$\Gamma_{\xi}$.

\subsection{Gaussian prior}
\begin{figure}[htb]
\centering
    \includegraphics[width=.48\textwidth]{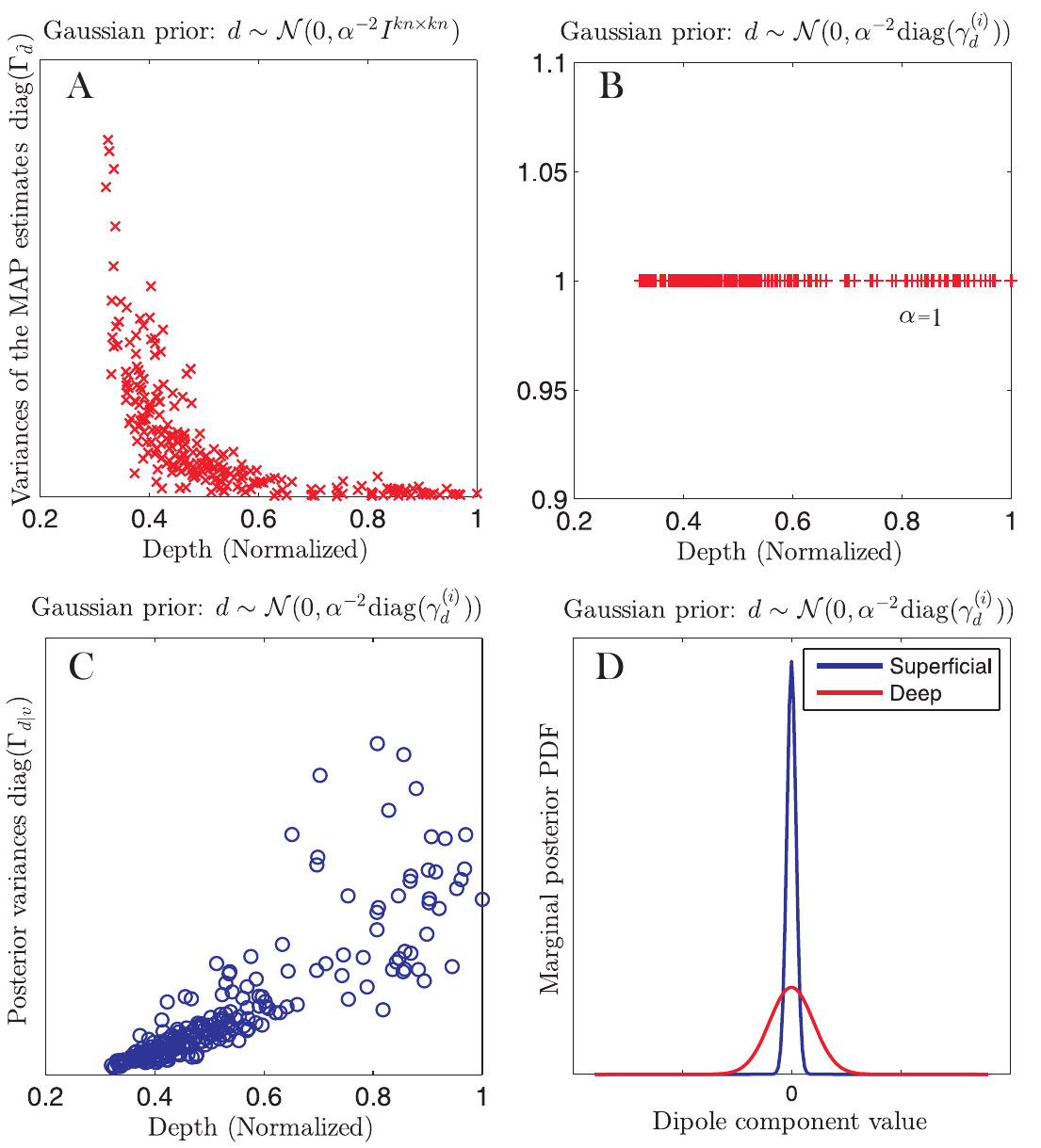}
\caption{A and B: The diagonal elements of $\Gamma_{\hat{d}}$ with
respect to depth when the i.i.d. and depth compensated Gaussian
prior are used, respectively. C: The posterior variances w. r. t.
depth when the depth compensated prior is used. D: Two marginal
distributions of the posterior.
%
%
}
\label{fig:MNEposteriorVariances}
\end{figure}

Let us consider a Gaussian prior
\begin{equation}
\pi(d)\propto\exp{\left(-\frac{1}{2}d^\mathrm{T}\Gamma_{d}^{-1}d\right)}
\end{equation}
 that does not have depth weights i.e. the covariance matrix is $\Gamma_{d}  = \alpha^{-2} I$ where
$I$ is the identity matrix and $\alpha^2$ a scaling parameter. In
this case, the MAP estimate is \cite{Hamalainen1994}
\begin{equation}\label{eq:Unweighted_MAP}
\hat{d} = K^\mathrm{T}(KK^\mathrm{T}+\alpha^2 \Gamma_{\xi})^{-1}v.
\end{equation}
From variational point of view, this MAP estimate coincides with
Tikhonov regularization and thus yields to a harmonic solution that
attains its maximum at the boundary \cite{Burger2013a}. This can
also be explained statistically by analyzing the expectation value
and covariance of the MAP estimates. Theses values can be estimated
by sampling or by using the analytical expressions
\begin{equation}\label{eq:CovUnweighted_MAP}
\mathbb{E}[\hat{d}]=0\;\;\; \mbox{ and }\;\;\;\Gamma_{\hat{d}} =
\mathbb{E}[\hat{d}\hat{d}^\mathrm{T}]=
K^\mathrm{T}(KK^\mathrm{T}+\alpha^2 \Gamma_{\xi})^{-1} K.
\end{equation}
Figure~\ref{fig:MNEposteriorVariances}-A shows how the values of the
diagonal elements (variances) of $\Gamma_{\hat{d}}$ decrease almost
quadratically with respect to depth. The zero expectation values and
the very low variances associated with the deep locations imply that
the deep sources are very unlikely to be reconstructed. Thus, this
MAP estimator is biased with respect to depth and favors sources
close to the sensors.

In this paper, our aim is to determine such prior variances that the
resulting MAP estimates do not favor any particular source location
or component over other i.e. the variances of the MAP estimates are
equal.

We start by postulating that this prior covariance matrix is
diagonal $\Gamma_{d} = \alpha^{-2}\mathrm{diag}(\gamma_d^{(i)}) $
for $i=1,\ldots,kn$. The MAP estimate corresponding to this prior is
\begin{equation}\label{eq:weighted_MAP}
\hat{d} = \Gamma_dK^\mathrm{T}(K\Gamma_dK^\mathrm{T}+\alpha^2
\Gamma_{\xi})^{-1}v,
\end{equation}
and the covariance of the MAP estimates becomes
\begin{equation}
\Gamma_{\hat{d}} = \mathbb{E}[\hat{d}\hat{d}^\mathrm{T}]=
\Gamma_dK^\mathrm{T}(K\Gamma_dK^\mathrm{T}+
\Gamma_{\xi})^{-1}K\Gamma_d.
\end{equation}
In a similar way as in \cite{eLORETA}, we estimate the prior
variances by minimizing
\begin{equation}
\gamma_d :=
\min_{\gamma_d}\|\mathrm{diag}(\alpha^{-2}I-\Gamma_{\hat{d}})\|_2^2.
\end{equation}
This results in solving a set of non-linear equations
\begin{equation}\label{}
\alpha^{2} =
\gamma_d^{2(i)}{K}^{(:,i)\mathrm{T}}M{K}^{(:,i)}\;\;\;\;\;\mbox{ for
} i=1,\ldots, kn,
\end{equation}
where $M=(\Gamma_{{\xi}}+K\Gamma_{{d}}K^\mathrm{T})^{-1}$ and
${K}^{(:,i)}$ is the $i^{th}$ column.

Figure \ref{fig:MNEposteriorVariances}-B shows that with these prior
variances the diagonal elements of $\Gamma_{\hat{d}}$ will be equal,
or in other words, the corresponding MAP estimator is depth
unbiased. Moreover, Figure \ref{fig:MNEposteriorVariances}-C depicts
the diagonal elements of the posterior covariance
$\Gamma_{d|v}=(K^\mathrm{T}\Gamma_{\xi}^{-1}K+\Gamma_{d}^{-1})^{-1}$
obtained based on the estimated prior
 and Figure
\ref{fig:MNEposteriorVariances}-D shows two corresponding marginal
posterior densities of two different locations. We can observe that
the posterior dipole variances increase with respect to depth.
Qualitatively, this means that in the estimated source
configurations the deep sources are allowed to have higher strengths
than the superficial sources, and therefore, the solutions can
attain their maximum also deeper in the brain (and not only close to
the sensors).

\subsection{$\ell_1/\ell_2$- norm sparsity prior} \label{prior} In this paper, we
consider sparse focal source reconstructions and therefore, we
employ the $\ell1/\ell_2$-norm prior 
\begin{equation}
\pi(d)\propto\exp{\left({-\frac{\alpha}2\sum_{i=1}^{n}w^r_i\|\mathrm{d}_{i}\|_2}\right)}\end{equation}
 where $d_{i}=(d_{ix},d_{iy},d_{iz})$,
$\|d_{i}\|_2=\sqrt{d_{ix}^2+d_{iy}^2+d_{iz}^2}$  is the strength of
the source at location $i$ and $w_i^r$ are the weights. 
For short, we denote the dipole strength at location $i$ as
${r}_i=\|\mathrm{d}_{i}\|_2$ and
\begin{equation}
\pi(r_{i})\propto \exp{\left(-\frac{\alpha}2 w^r_ir_i\right)}.
\end{equation}
The variance of $\pi(r_i)$ is
\begin{equation}\label{eq:VarianceDefinition}
\gamma_{r}^{(i)}=
c\int_0^\infty (r_i-r_{*i})^2\exp{\left(-\frac{\alpha}2 w_{i}^r r_i\right)}~dr_i
=\frac{4}{\alpha^2{(w^r_{i})}^2}
\end{equation}
where $r_{*i} =c\int_0^\infty r_i\exp{\left(-0.5\alpha
w_i^rr_i\right)}~dr_i=\frac{4c}{\alpha^2 {(w_i^r)}^2}$ and
$c=0.5\alpha w_i^r$ because $\int_0^\infty c\exp{\left(-0.5\alpha
w_{i}^r r_i\right)}~dr_i = 1$.

\begin{figure}[htb] \centering
    \includegraphics[width=.47\textwidth]{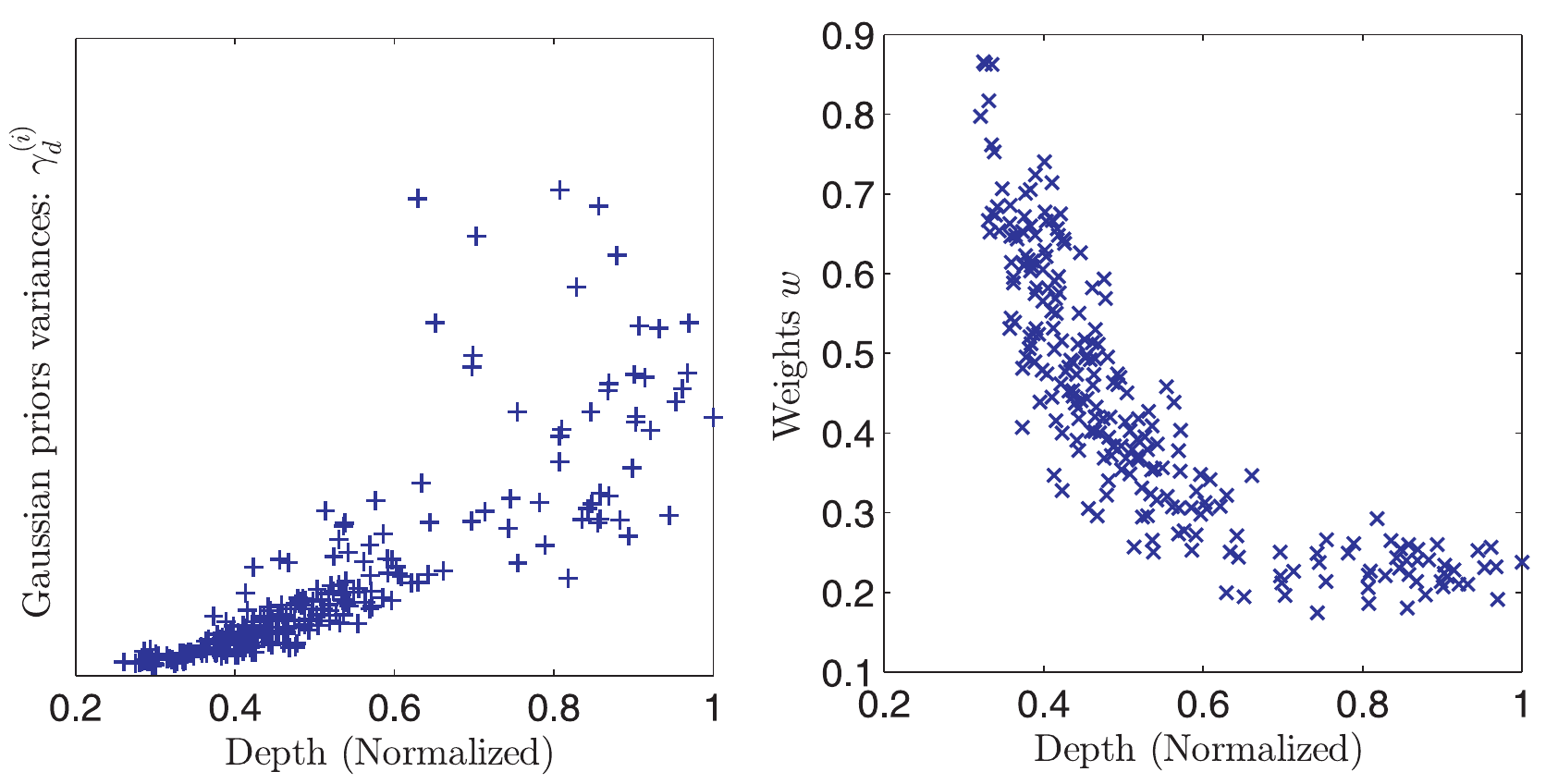}
\caption{The estimated Gaussian prior variances $\gamma_{d}^{(i)}$
and the corresponding weights for the
$\ell_1/\ell_2$ norm prior with respect to depth 
.
}\label{fig:L12_weights}
\end{figure}

We calculate $\gamma_{r}^{(i)}$ at location $i$ with the help of the
corresponding Gaussian variances 
$\gamma_d^{(i+(j-1)n)}$ as
\begin{equation}\label{eq:VarianceSelection}
\gamma_{r}^{(i)}=k\alpha^{-2k+2}\left(\prod\limits^{k}_{j=1}\gamma_d^{(i+(j-1)n)}\right)
\left(\sum\limits^{k}_{j=1}\gamma_d^{(i+(j-1)n)}\right)^{-1},
\end{equation}
where  $j=1,\ldots,k$ and $k$ is the dimension of the problem. This
choice ensures that $\gamma_{r}^{(i)}$ is roughly the average of the
dipole component variances when the variances of the components are
similar and that $\gamma_{r}^{(i)}$ is close to the lowest dipole
component variance when the variances have large differences.
Finally, from Equation (\ref{eq:VarianceDefinition}) and
(\ref{eq:VarianceSelection}) we calculate the weights
\begin{equation}\label{eq:proposedweights}
w_i^r=2\sqrt{\frac{\alpha^{2k-4}}{k}\frac{\sum\limits_{j=1}^{k}\gamma_d^{(i+(j-1)n)}}{2\prod\limits_{j=1}^{k}\gamma_d^{(i+(j-1)n)}}}
\end{equation}
The estimated Gaussian variances and the corresponding weights of
the $\ell_1/\ell_2$-norm prior are shown in Figure
\ref{fig:L12_weights}.
\section{Materials and methods}
We study the proposed weights by simulating focal deep sources in
the gray matter of a 2D FE head model. The head model consisted of
five compartments with conductivities (in S/m) equal to $0.33$ for
the scalp, $0.015$ for the skull, $1.76/0.016/0.33$ for the cerebral
spinal fluid, gray matter and white matter \cite{Vorwerk2014},
respectively. The potential measurements $v$ were obtained from 32
point sensors equally spaced around the boundary. For the forward
and the inverse computations, we use two meshes with 2342 and 1236
nodes, respectively.

The MAP estimate of the dipole configuration with sparsity
constraint is
\begin{equation}\label{eq:L2_norm}
\hat{d}_\mathrm{MAP} :=\min_d\|v-Kd\|_2^2+\sum_{i=1}^{n}\lambda
w_i\|\mathrm{d}_{i}\|_2 w
\end{equation}
where $\lambda$ is a tuning parameter. The minimization is performed
by using the interior point method \cite{Boyd2004} with Bregman
iterations \cite{Yin2008}. The performance of the proposed weights,
$w_{i}^r$, from Equation (\ref{eq:proposedweights}), was compared
with two other commonly used weights: first, the MNE resolution
weights given by $w_i^\mathrm{MNE} = \sqrt{1/k
\sum_{j=1}^kR^{(i,i+(j-1)n)}}$, where $R^{(i,i)}=\mathrm
{diag}(K^\mathrm {T} (KK^\mathrm {T}+\Gamma_{\xi})^{-1}K)$
\cite{Haufe2008b} and second, the normalized maximum sensor
responses $w^{\mathrm MSR}_i=g_i/\max{(g_i)}$, where
$g_i=\max_{l=1:m}\left( \| 1/k\sum_{j=1}^k K^{(l,i+(j-1)n)}
\|_2\right)$ \cite{Fuch1994}. To access the ground truth, we
consider measurements with high signal to noise ratio,
$\mathrm{SNR}=60$dB. For the quantitative comparison of the results
we employ the earth mover's distance (EMD) \cite{Rubner2000}.

\section{Results and discussion}

\begin{figure}[!htb]
\centering
  \includegraphics[width=0.452\textwidth]{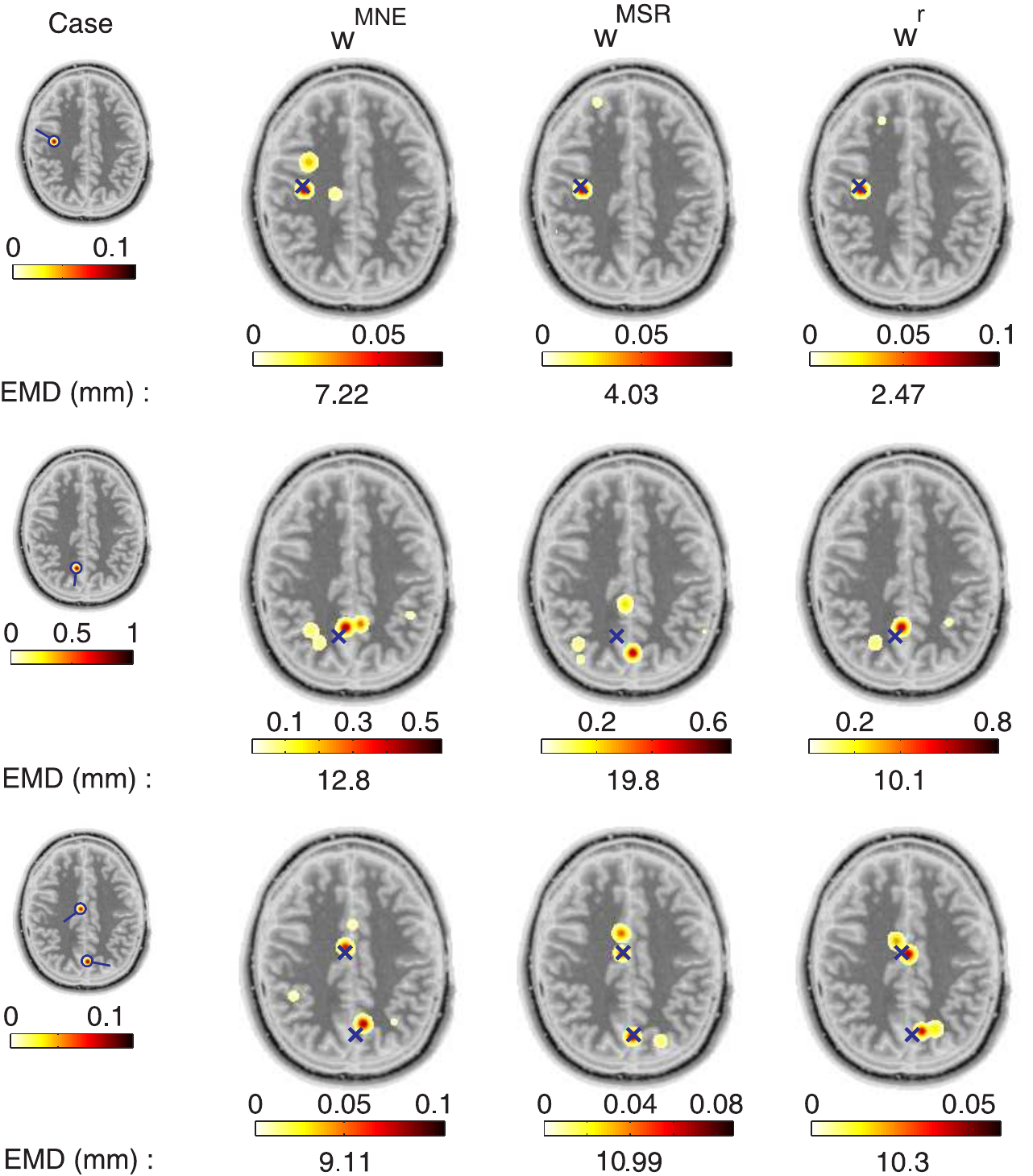}
\caption{Reconstructed source distributions using different weights
in the $\ell_1/\ell_2$ prior model. The images show first the test
cases and then the reconstructions with the different weights
$w_i^\mathrm{MNE}$, $w_i^\mathrm{MSR}$ and $w_i^{r}$, respectively.}
\label{fig:mri_1s}
\end{figure}

We demonstrate the performance of the different weights using three
test cases with one and two dipole sources. In
Figure~\ref{fig:mri_1s}, the small images on the left hand side show
the true dipoles, the location is marked with blue circles and the
orientations with small blue lines. The remaining images, starting
from left, show the reconstruction when $w_i^{\mathrm{MNE}}$,
$w_i^\mathrm{MSR}$ and $w_i^{r}$ are used as weights, respectively.
The blue marker $\mathrm{x}$ shows the locations of true sources.
The MAP estimates were computed by solving Equation
(\ref{eq:L2_norm}).

All the tested weights give feasible reconstructions. However, we
note that the proposed weights $w_i^r$ give the least scattered
results and work the best in the single focal source cases. For the
two source case, all the weights give roughly similar
reconstructions and EMD values.

\section{Conclusion and future work}

We have demonstrated that the proposed depth weights with the
$\ell_1/\ell_2$ sparsity prior give better reconstruction compared
to two commonly used weights when single deep sources are studied.
Our proposed approach has the benefit that it does not require using
hyper-parameter models that would involve extensive sampling due to
the lack of an analytical expression for the MAP estimate when the
$\ell_1/\ell_2$ prior is used. In the future, Monte Carlo
simulations will be carried out in a 3D head model to analyze the
distribution of the MAP estimates reconstructed by using the
$\ell_1/\ell_2$ sparsity prior with the proposed weights.

\section*{Conflict of interest}

The authors declare that they have no conflict of interest.

\bibliography{koulouri}

\begin{table}[h]
\footnotesize
        \begin{tabular}{ll}
        &Corresponding author: Alexandra Koulouri \\
        &Institute: University of M\"unster\\
        &Street: Einsteinstrasse 62\\
        &City: M\"unster\\
        &Country:  Germany\\
        &Email: koulouri@uni-muenster.de \\
        \end{tabular}
\end{table}

\end{document}